\documentclass[prl,preprint,superscriptaddress]{revtex4-2}
\usepackage{dcolumn}
\usepackage{mathtools}
\usepackage{amssymb}
\usepackage{bm}
\usepackage{pifont}
\usepackage{psfrag}
\usepackage{epstopdf}
\usepackage{wasysym}
\usepackage{amsmath}
\usepackage{tipa}
\usepackage{hyperref}
\usepackage{lineno}
\usepackage{lipsum}
\usepackage[normalem]{ulem}
\usepackage{graphicx}
\usepackage{setspace}
\hypersetup{
     colorlinks = true,
     linkcolor = blue,
     anchorcolor = blue,
     citecolor = blue,
     filecolor = blue,
     urlcolor = blue
     }
\usepackage[usenames]{color}
\renewcommand{\footnoterule}{%
  \hrule width \textwidth height 1pt
  \kern 2pt
}

\begin{document}
\title{A shape-driven reentrant jamming transition in confluent monolayers of synthetic cell-mimics}

\author{Pragya Arora}
\affiliation{Chemistry and Physics of Materials Unit, Jawaharlal Nehru Centre for Advanced Scientific Research, Jakkur, Bangalore - 560064, INDIA}
\author{Souvik Sadhukhan}
\affiliation{Tata Institute of Fundamental Research, Hyderabad - 500046, INDIA}
\author{Saroj Kumar Nandi}
\affiliation{Tata Institute of Fundamental Research, Hyderabad - 500046, INDIA}
\author{Dapeng Bi}
\affiliation{Department of Physics, Northeastern University, Boston, MA 02115, U.S.A.}
\author{A K Sood}
\affiliation{Department of Physics, Indian Institute of Science, Bangalore- 560012, INDIA}
\affiliation{International Centre for Materials Science, Jawaharlal Nehru Centre for Advanced Scientific Research, Jakkur, Bangalore - 560064, INDIA}
\author{Rajesh Ganapathy}
\affiliation{International Centre for Materials Science, Jawaharlal Nehru Centre for Advanced Scientific Research, Jakkur, Bangalore - 560064, INDIA}
\affiliation{School of Advanced Materials (SAMat), Jawaharlal Nehru Centre for Advanced Scientific Research, Jakkur, Bangalore - 560064, INDIA}

\date{\today}
\begin{abstract}
\textbf{Many critical biological processes, like wound healing, require confluent cell monolayers/bulk tissues to transition from a jammed solid-like to a fluid-like state. Although numerical studies anticipate changes in the cell shape alone can lead to unjamming, experimental support for this prediction is not definitive because, in living systems, fluidization due to density changes cannot be ruled out. Additionally, a cell's ability to modulate its motility only compounds difficulties since even in assemblies of rigid active particles, changing the nature of self-propulsion has non-trivial effects on the dynamics. Here, we design and assemble a monolayer of synthetic cell-mimics and examine their collective behaviour. By systematically increasing the persistence time of self-propulsion, we discovered a cell shape-driven, density-independent, re-entrant jamming transition. Notably, we observed cell shape and shape variability were mutually constrained in the confluent limit and followed the same universal scaling as that observed in confluent epithelia. Dynamical heterogeneities, however, did not conform to this scaling, with the fast cells showing suppressed shape variability, which our simulations revealed is due to a transient confinement effect of these cells by their slower neighbors. Our experiments unequivocally establish a morphodynamic link, demonstrating that geometric constraints alone can dictate epithelial jamming/unjamming.
}
\end{abstract}
\maketitle

Confluent cell monolayers and tissues are a maximally crowded environment; the cell packing fraction is nearly unity. And yet, remodeling and repair occur, and cancer cells migrate from a tumor and invade distal sites. These processes require tissue/cell collectives to flow locally \cite{trepat2009natphys,angelini2011glass,sadati2013collective,schoetz2013glassy,garcia2015physics,kas2017jphysd,atia2021celldev,kas2021frontiers}, and the jamming-unjamming transition, akin to the one seen in inert particle assemblies \cite{liu1998nature}, provides a pathway. In inert particle assemblies, the unjamming transition is density-driven \cite{liu2010annrev,behringer2018repphys}, and there is evidence of such a transition during embryonic morphogenesis \cite{mongera2018nature} and cancer invasion \cite{han2020natphys,ilina2020natcellbio,grosser2021prx}. But cells, unlike inert particles, can deform to surmount the constraining effects of crowding \cite{kas2017jphysd,atia2021celldev,kas2021frontiers,stephens2021physics}. In fact, in the vertex model of confluent epithelia \cite{nagai2001philmagb,farhadifar2007currbio}, the competing effects of cell contractility and cell-cell adhesion lead to a density-independent but cell shape change driven unjamming transition  \cite{bi2015density}. In the jammed state, the cells have a more regular shape than in the fluid state, where they are more elongated. A subsequent model that included cell motility showed the jamming transition could be driven by additional factors, but the qualitative shape-based nature remained unchanged \cite{bi2016jamming}. Cell shape is proving to be a structural marker of unjamming \cite{kas2017jphysd,atia2021celldev,kas2021frontiers,stephens2021physics,trepat2018mesoscale}, and this is the case even when alignment interactions between neighboring cells are present \cite{malinvero2017natmat,giavazzi2018softmatter,wang2020pnas,paoluzzi2021pre}. Cell shape-mediated unjamming is now implicated in the pathophysiology of asthma \cite{park2015unjamming} and tumor progression \cite{grosser2021prx,gottheil2023prx}.

Besides shape, cells in a collective also show substantial shape variability, which, until recently, was dismissed as noise. Atia et al., \cite{atia2018geometric}, observed that across vastly different epithelial systems, the cell shape distribution became progressively less skewed as the system jammed. When scaled appropriately, these distributions collapsed to a universal distribution known as the $k-$gamma distribution \cite{aste2008pre}. Remarkably, across these different systems, the shape variability and the cell shape follow a simple linear relationship; the larger the cell aspect ratio, the greater the shape variability. These results suggest that, like in inert particle packings \cite{aste2008pre,boromand2018prl}, geometric constraints take center stage in tissue jamming/unjamming.

Tissues, however, are complex, and besides cell shape changes, jamming-unjamming can be driven by cell division, apoptosis, extrusion, cell size changes, and motile topological defects \cite{saw2017nature}. These processes may operate independently or work in cohorts \cite{atia2018geometric}, and being impossible to suppress, there is no consensus if cell shape changes alone drive unjamming \cite{kas2017jphysd,kas2021frontiers}. Additionally, cells can regulate their motility, which is another critical parameter governing glass/jamming physics \cite{berthier2017njp,nandi2018pnas}. Even for the simpler case of rigid self-propelled particles, tuning the activity at constant density results in non-intuitive behavior, such as re-entrant dynamics \cite{berthier2014prl,klongvessa2019prl,arora2022prl}, and it is unknown whether similar physics is at play in cell collectives, since, here, the means to tune activity systematically are not available. Synthetic models that embody the key features of confluent epithelia can help bridge the divide between theory/numerical predictions and experiments on living systems. They also have the potential to single out the role of geometric constraints in tissue jamming/unjamming. In fact, a recent theory inspired by \cite{atia2018geometric} posits that the relationship between shape variability and cell shape is purely a mathematical property of a confluent monolayer of closed-loop objects \cite{sadhukhan2022origin}, but this idea still needs to be tested in a synthetic model system.

\subsection*{Making deformable cell-mimics with tunable activity} 
Here, we create synthetic active cell-mimics, assemble them into monolayers, and examine their collective behavior. At the single-cell level, our mimics possess just two features: deformability and tunable activity. Existing deformable active matter systems - centimeter-sized bots confined within paper/metal rings \cite{boudet2021scirob} and active colloids confined within vesicles \cite{vutukuri2020nature} do not serve our purpose. In the former system, each cell is itself large ($\approx 30$ cm in diameter), making it a challenge to study collective behavior; with the latter system, methods to create assemblies of these and regulate their activity are unavailable. 

\begin{figure}[tbp]
\includegraphics[width=1\textwidth]{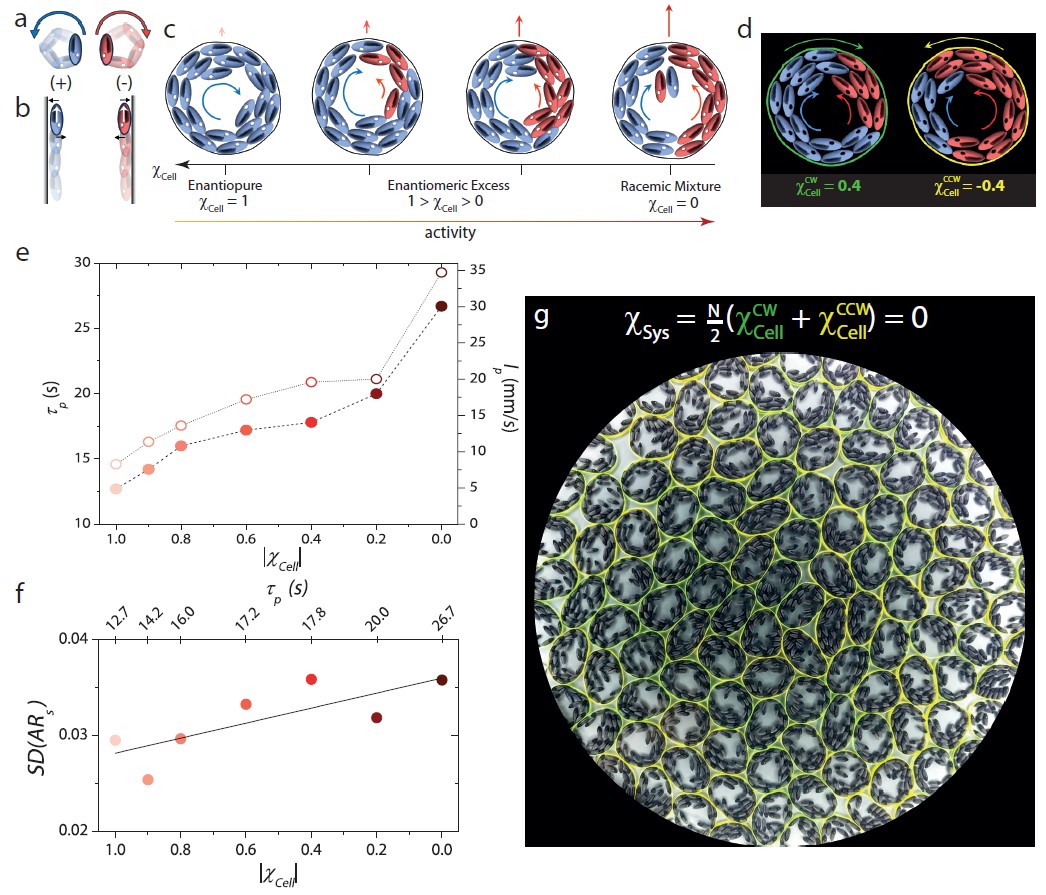}
\caption{\textbf{Making a confluent layer of synthetic cell mimics.} \textbf{(a)} Superimposed snapshots showing a nearly circular path traced by polar chiral active ellipsoids under vertical agitation.  Clockwise $(+)$ and counterclockwise $(-)$ moving ellipsoids are represented in blue and red colours, respectively. 
\textbf{(b)} Polar chiral active ellipsoids align with their propulsion axis parallel to the wall with the direction of motion along the wall determined by the ellipsoid's handedness.
\textbf{(c)} Granular cell enclosing $N=20$ polar chiral active ellipsoids. Panels from left to right show cells with the chirality of the cell interior, $\chi_{\text{Cell}}$, varying from enantiopure to racemic. While a unidirectional polarized wall current results in cell spin, counter-propagating currents result in cell motility, the extent of which can be tuned by systematically decreasing $|\chi_{\text{Cell}}|$. 
\textbf{(d)} The handedness of cell spin is determined by the sign of $\chi_{\text{cell}}$. The green and yellow membranes spin in clockwise and anticlockwise directions due to an excess of $(+)$ ellipsoids in one and $(-)$ ellipsoids in the other.
\textbf{(e)} Persistence time $\tau_{p}$ (hollow circles) and the persistence length $l_{p}$ (solid circles) versus $|\chi_{\text{cell}}|$ of isolated cells. \textbf{(f)} Shape variability of isolated cells, $SD(AR_s)$, versus $|\chi_{\text{cell}}|$. The greater the cell activity, the greater the shape variability. \textbf{(g)} Snapshot of a nearly confluent assembly ($\phi\approx0.94$). The net chirality of the system, $\chi_{\text{Sys}} = 0$. The plate has an equal number of clockwise (green) and counterclockwise (yellow) spinning cells. 
}
\label{Figure1}
\end{figure}

Our cell-mimics instead are flexible paper rings ($3$ cm in diameter) that enclose a monolayer of 3D-printed granular ellipsoids rendered chiral active by vertical vibration. These ellipsoids experience both an active torque and force under such a driving and perform circle active motion with the handedness of the trajectory, i.e., clockwise $(+)$ or counterclockwise $(-)$, manually determined during particle placement on our shaker apparatus \cite{arora2021sciadv} (Fig. \ref{Figure1}a top panel \& see Materials and Methods and Fig. S1-S4). Our preliminary observations guided us to leverage chiral activity rather than achiral activity to make our cells self-propelled. On confining achiral polar active ellipsoids within the paper rings, the interplay of membrane curvature and particle orientation often resulted in their accumulation at diametrically opposite ends with their polarity pointing outwards \cite{boudet2021scirob,paoluzzi2016scirep,peterson2021natcomm}. Besides resulting in low cell motility, this preferential accumulation led to a nonuniform cell stiffness (Supplementary Fig. S4 and Supplementary Movie 1). In contrast, when persistent active torques are also present, particles hug a wall with their propulsion axis parallel to it and with the direction of motion along the wall set by the handedness of the activity (Fig. \ref{Figure1}a bottom panel) \cite{barois2020sorting}. Supplementary Movie 2 shows the dynamics of a granular cell with $N=20$ chiral ellipsoids of the same handedness, $(-)$, within (Figure \ref{Figure1}c(i)). Unlike with the achiral ellipsoids, the particles now uniformly decorated the membrane interior with the polarized particle current acting like a ``dynamic internal skeleton". However, this current only caused the paper ring to spin with a well-defined handedness, but we observed little translation. 

To tune the motility of the cell-mimics, we changed the handedness of a few ellipsoids inside the membrane from $(-)$ to $(+)$ (red ellipsoids in Fig. \ref{Figure1}c(ii)-(iv)). We observed counter-propagating particle currents that suppressed cell spin and made it motile (see Supplementary Movie 2 and Supplementary Table I). Thus, even when $N = N_{+} + N_{-}$ is held fixed at twenty particles, changing the magnitude of chirality of the cell interior, $|\chi_{\text{cell}}|$, helps tune cell activity. Here, $\chi_{\text{cell}} = \frac{N_{+} - N_{-}}{N_{+} + N_{-}}$,  and $N_{+}$ and $N_{-}$ are the number of $(+)$ and $(-)$ ellipsoids, respectively. We note that the handedness of cell spin depends on the sign of $\chi_{\text{cell}}$, but the internally generated active force depends only on its magnitude. For example, the green and yellow cells depicted in Fig. \ref{Figure1}d spin in opposite directions due to an excess of $(-)$ ellipsoids in one and $(+)$ ellipsoids in the other; however, the active force is determined solely by $|N_{+} - N_{-}|$, which is the same for both.

We quantified the activity of isolated cells for different $|\chi_{\text{cell}}|$ values by working in the low area fraction limit of cells, $\phi<1\%$. We measured the persistence time, $\tau_p$ - the time at which the cells' mean-squared displacement crossed over from ballistic to diffusive dynamics (Supplementary Fig. S5a), and also the average cell speed, $v$ (Supplementary Fig. S5b) \cite{arora2022prl}. A larger value of $\tau_p$ signifies a stronger departure from equilibrium, i.e., \textit{greater} activity. When the cell interior was gradually changed from enantiopure ($|\chi_{\text{cell}}| = 1)$ to racemic ($|\chi_{\text{cell}}| = 0)$, both $\tau_p$ and the persistence length, $l_p = v\tau_p$, increased in a systematic manner (Fig. \ref{Figure1}e, Supplementary Table II). Recent studies have identified $\tau_p$ as a crucial parameter governing active glass physics \cite{berthier2014prl,nandi2018pnas,henkes2020natcomm,arora2022prl}, and our approach allows for systematically tuning it. Interestingly, the active torques and forces exerted by the particles on the membrane interior also influenced the shape variability of isolated cells. We used the standard deviation, $SD$, of an isolated cells' aspect ratio, $AR_s$, as a measure of shape variability, and this is shown in Fig. \ref{Figure1}f for different $|\chi_{\text{cell}}|$ values. In the limit of an enantiopure cell interior, the persistent torques due to the unidirectional particle current at the boundary suppressed cell shape fluctuations, resulting in small $SD$, while on approaching a racemic cell interior, the more frequent reorganization of the counter-propagating particle currents made it floppy (large $SD$, see Supplementary Movie 2).

\subsection*{Cell-mimic collective}
Before we can ascertain if our synthetic cells in the dense limit embody the crucial features of confluent epithelia, another parameter, namely, the net chirality, $\chi_{\text{Sys}}$, of the synthetic cell assemblies must be adjusted. Here, $\chi_{\text{Sys}}= \frac{|N_{\text{CW}}-N_{\text{CCW}}|}{N_{\text{CW}}+N_{\text{CCW}}}$, where $N_{\text{CW}}$ and $N_{\text{CCW}}$ are the number of clockwise and counterclockwise spinning cells. We observed that $\chi_{\text{Sys}}$ had a qualitative effect on the dynamics; for example, in a nearly confluent assembly ($\phi\approx0.94$) of clockwise spinning cells, i.e., $\chi_{\text{Sys}} = 1$ (green cell in Fig. \ref{Figure1}d), there was an emergent edge current (Supplementary Movie 3) like those seen in other confined chiral active matter systems \cite{van2016spatiotemporal,beppu2021edge}.
However, except in very rare cases and specific cell lines \cite{duclos2018natphys,yashunsky2022prx},
most experiments examining the jamming-unjamming of confluent epithelia do not report such edge currents \cite{trepat2009natphys,angelini2011glass,park2015unjamming,atia2018geometric}, suggesting that these assemblies do not possess an overall chirality. To make contact with these studies, we, therefore, set $\chi_{\text{Sys}} = 0$ by having an equal number of clockwise (green) and counterclockwise (yellow) spinning cells on the plate for all values of $\phi$ and $\tau_p$ studied (see Fig. \ref{Figure1}f for $\phi = 0.92$). The edge flows were absent in these assemblies. (Supplementary Movie 4). 

\subsection*{A re-entrant jamming transition mediated by cell shape}
\begin{figure}[tbp]
\includegraphics[width=1\textwidth]{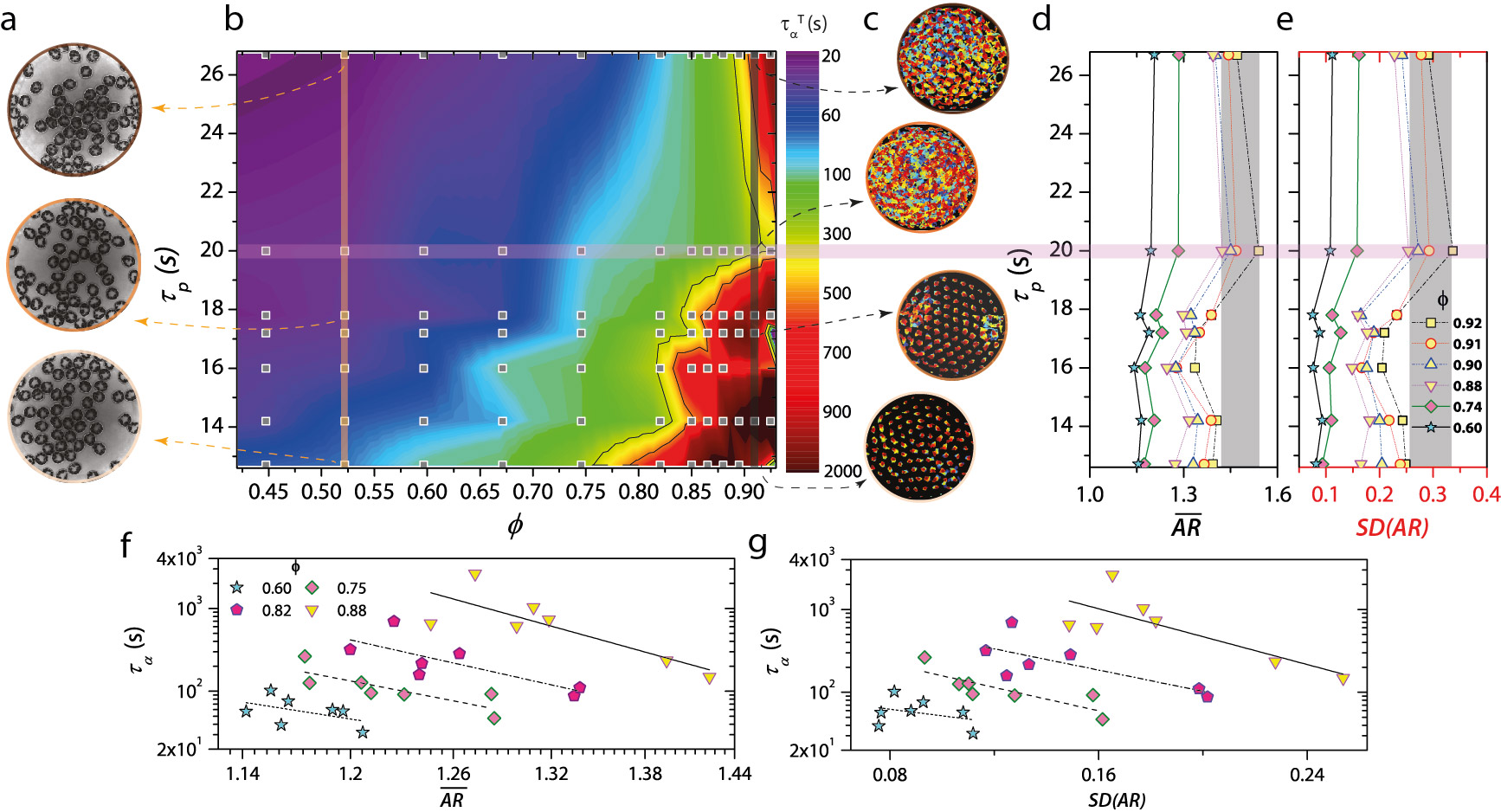}
\caption{\textbf{Cell shape and shape variability are correlated with dynamics for a confluent cell monolayer.} 
\textbf{(a)} Snapshots of the granular cell assemblies at $\phi=0.44$ for three representative activities. Note a weak tendency of the cells to form clusters with increasing $\tau_p$.
\textbf{(b)} Relaxation dynamics phase diagram in the ($\phi$, $\tau_p$) plane. The squares represent the $\chi_{\text{Cell}}$ and $\phi$ at which experiments were performed. The black dashed lines are the isochrones. The colour bar indicates the value of the relaxation time $\tau_{\alpha}$. The $\tau_{\alpha}$ values between experimental data points were obtained by linear interpolation.
\textbf{(d) \& (e)} $\overline{AR}$ and $SD(AR)$ for different $\tau_p$ values are shown for a range of $\phi$. The window of $\phi$ values where the re-entrant behavior was seen is represented by yellow-filled symbols. Both $\overline{AR}$ and $SD(AR)$ are largest at the intermediate value of $\tau_p$ (horizontal bar), where the system exhibits fluid-like behavior.
\textbf{(f) \& (g)} $\tau_{\alpha}$ versus $\overline{AR}$, and $\tau_{\alpha}$ versus $SD(AR)$ for different values of $\phi$. 
}
\label{Figure2}
\end{figure}

Next, by analyzing the dynamical trajectories of the cell centers (see Materials and Methods), we charted the relaxation dynamics of the granular cell assemblies for different values of $\phi$ and $\tau_p$ (Fig. \ref{Figure2}a-c). We determined the structural relaxation time, $\tau_\alpha$, as the time at which the self-intermediate scattering function $F_{s}(q, t)$ decayed to $\frac{1}{e}$ (Supplementary Fig. S6). Here, $q$ is the wavevector and was chosen to correspond to the inverse of the cell diameter. While for a fixed value of $\tau_p$, increasing $\phi$ results in dynamical slowing down, as expected, the effects of increasing $\tau_p$ at fixed $\phi$ is more subtle (Fig. \ref{Figure2}b). For $0.4\leq \phi \leq0.88$, an increase in $\tau_p$ sped up the dynamics, and over a narrower window $0.4\leq \phi \leq0.7$, we observed a weak tendency of the cells to form transient clusters (Fig. \ref{Figure2}a and Supplementary Movie 5) This clustering without attractive interactions is a cardinal feature of active matter \cite{fily2012prl,cates2015annrev} and was evident as a shoulder in the distribution of Voronoi areas (Supplementary Fig. S7). Notably, over a range of densities close to confluence ($0.88 \leq \phi \leq 0.92$), increasing $\tau_p$ resulted in a re-entrant behavior: structural relaxation was fastest at an intermediate value of $\tau_p$ (horizontal bar in Fig. \ref{Figure2}b). This re-entrant behavior is evident in the cell trajectories, with the cells being strongly caged (glass-like) for both small and large values of $\tau_p$ and ergodic (fluid-like) for an intermediate value (vertical bar corresponding to $\phi \approx 0.91$ in Fig. \ref{Figure2}c, see Supplementary Movie 6).

On increasing the strength of particle attraction in dense assemblies of hard passive particles, the competing effects of particle crowding and attraction often lead to re-entrant dynamics \cite{pham2002science,mishra2013prl}. Similar behavior is seen in dense assemblies of hard active particles on increasing $\tau_p$ \cite{berthier2017njp,berthier2014prl,klongvessa2019prl,arora2022prl}, with the difference being that the attraction itself is activity-mediated, and its strength is proportional to $\tau_p$. However, unlike in hard passive/active particle systems, where the re-entrant behavior begins to manifest even at moderate densities \cite{mishra2013prl,berthier2014prl,arora2022prl}, we found it only on nearing confluence, which suggests that cell shape changes may have a vital role here. We calculated the average cell aspect ratio $\overline{AR}$ of the assemblies for a range of $\phi$ and all values of $\tau_p$ studied (Fig. \ref{Figure2}d). Remarkably, over the window of $\phi$ values where we observed a re-entrant behavior (yellow-filled symbols in Fig. \ref{Figure2}d), it is at the intermediate value of $\tau_p$ where the system was fluid-like that $\overline{AR}$ is largest (horizontal bar in Fig. \ref{Figure2}d). Even in our synthetic system, cell shape governs structural relaxation: assemblies of elongated cells relax faster (Fig. \ref{Figure2}f) \cite{grosser2021prx,bi2015density,bi2016jamming,park2015unjamming}. Indeed, for moderate densities ($\phi = 0.75\text{, } 0.60$), not only does the lack of re-entrant dynamics coincide with the absence of a non-monotonicity in $\overline{AR}$ with $\tau_p$ (Fig. \ref{Figure2}d), the correspondence between $\tau_\alpha$ and $\overline{AR}$ is also weaker (Fig. \ref{Figure2}f).

The above findings suggest that shape variability may also bear the imprint of the dynamics of these assemblies \cite{atia2018geometric}. We quantified the shape variability of our cell assemblies via the standard deviation of the aspect ratio $SD(AR)$. Strikingly, $SD(AR)$ mirrors the behavior of $\overline{AR}$ for different values of $\phi$ and $\tau_p$ (Fig. \ref{Figure2}e). Additionally, similar to $\overline{AR}$, there is a correlation between $SD(AR)$ and $\tau_\alpha$, with faster relaxation observed in assemblies with greater shape variability (Fig. \ref{Figure2}g). 

We can now explain the observed re-entrant behavior. For small $\tau_p$ values (large $|\chi_{\text{cell}}|$), besides the low cell motility, each cell also resists significant shape fluctuations (Fig. \ref{Figure1}f) and adopts a circular shape (small $\overline{AR}$). Since disk assemblies freeze into glasses at densities smaller than elongated particle ones as they require fewer constraints \cite{kas2021frontiers}, the system is glassy near confluence (Supplementary Movie 6). At the largest value of $\tau_p$ (i.e., $|\chi_{\text{cell}}| = 0$), individual cell motility is large, and the cells are also floppier (Fig. \ref{Figure1}f), but so is the activity-mediated adhesion between cells. Near confluence, this cell-cell adhesion brings down $\overline{AR}$ and $SD(AR)$, and the system is again glassy. However, these effects compete for an intermediate $\tau_p$ value: individual cell motility and shape variability are reasonably large, but adhesion is weak, making the assembly fluid-like. As far as we know, this is the first observation of a shape-mediated re-entrant jamming.

\subsection*{Shape and shape variability in confluent cell-mimic monolayers}

\begin{figure}[tbp]
\includegraphics[width=1\textwidth]{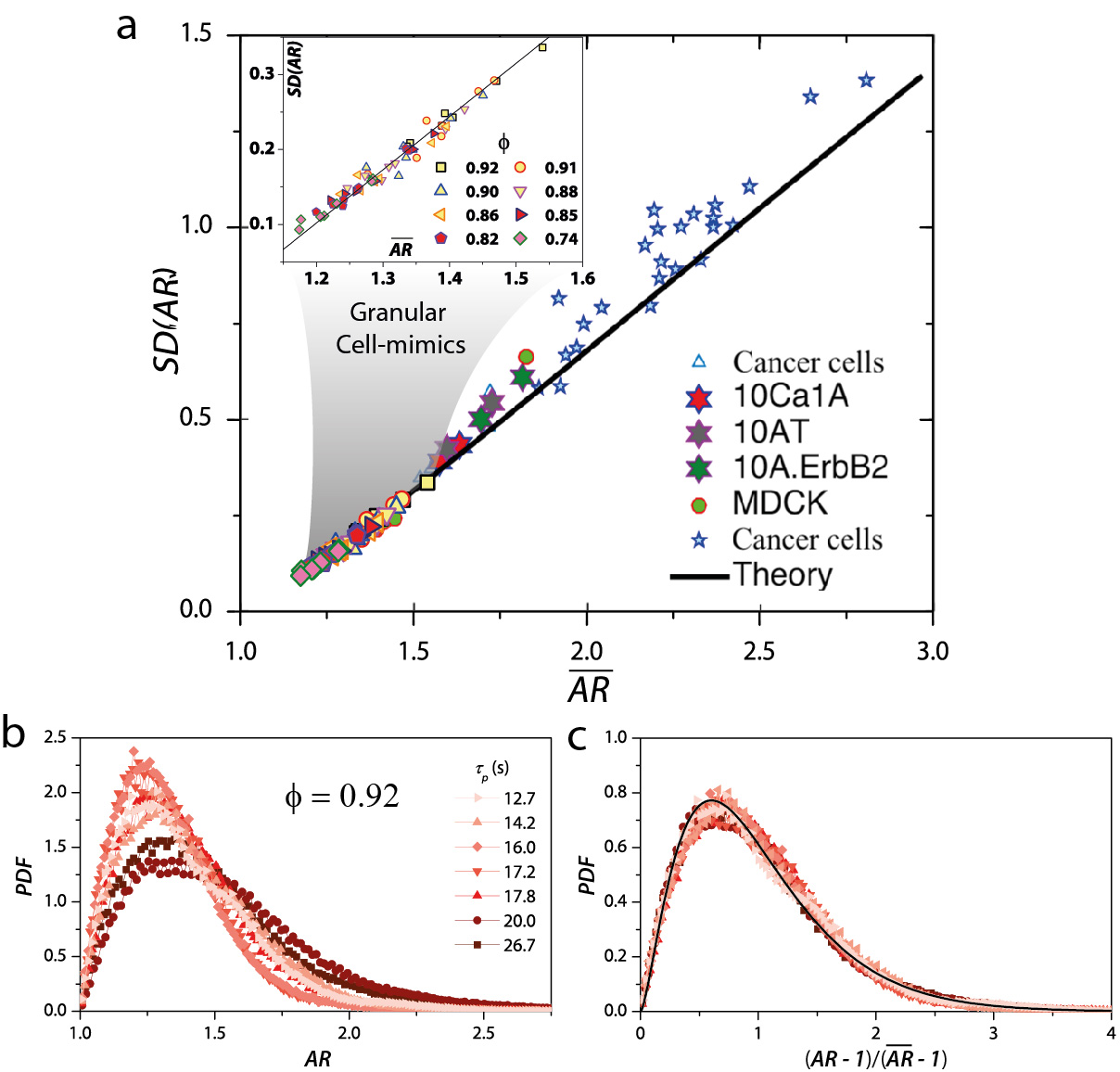}
\caption{\textbf{Cell shape and shape variability are mutually constrained for granular cell-mimics.} 
\textbf{(a)} $SD(AR)$ versus $\overline{AR}$ for granular cells adhere to the same universal scaling as seen in living systems. The black line is the theoretical prediction of ref. \cite{sadhukhan2022origin} and has the functional form ($SD(AR) \simeq 0.71\overline{AR}-0.75$).   
\textbf{(b)} Probability density functions (PDFs) of $AR$ 
for different $\tau_p$ values for $\phi = 0.92$. Note that $PDF$ exhibits a broad distribution with a large positive skew for the unjammed (fluid-like) state at an intermediate value of $\tau_p = 20$ s. For the more jammed states, corresponding to other $\tau_p$ values, the $PDF$s are narrower, and with a smaller skew. 
\textbf{(c)} When scaled by $(AR-1) /(\overline{AR}-1)$, the PDF at different $\tau_p$ values collapse. The black line represents a $k-$Gamma distribution, and the data can be equally well-fit to the distribution proposed in ref. \cite{sadhukhan2022origin}. 
}
\label{Figure3}
\end{figure}

The strikingly similar behavior of $\overline{AR}$ and $SD(AR)$ for different $\phi$ and $\tau_p$ values (Fig. \ref{Figure2}d \& e) suggests these two quantities are interdependent. Indeed, for our granular cells, $SD(AR)$ scales linearly with $\overline{AR}$ like that observed in vastly different epithelial systems (Fig. \ref{Figure3}a) \cite{atia2018geometric,sadhukhan2022origin}. We now checked whether this linear scaling was simply a result of the probability distribution function ($PDF$) of the aspect ratio being universal and a $k-$gamma distribution \cite{atia2018geometric,aste2008pre}. Figure \ref{Figure3}b shows the $PDF$ for different $\tau_p$ values for $\phi = 0.92$. The $PDF$ is broad with a large positive skew for the unjammed (fluid-like) state, which, here, is observed at an intermediate value of $\tau_p = 20$ s. For the more jammed states, corresponding to other $\tau_p$ values, the $PDF$s are narrower and with a smaller skew. This behavior is also seen for other values of $\phi$ near confluence (Supplementary Figs. S8-S11). Moreover, the $PDF$s at different $\tau_p$ values collapse onto a single universal distribution on rescaling $AR$ to $x = \frac{(AR-1)}{(\overline{AR}-1)}$, which ensures that the scaled $PDF$s start at the origin and have a mean of unity. Within experimental uncertainty, the scaled $PDF$s are equally well-fit by the $k-$gamma distribution \eqref{eq:1} with the value of $k \approx 2.51$ \cite{atia2018geometric}, as well as the recently posited distribution for a confluent assembly of closed-loop objects \eqref{eq:2}, which is only \textit{nearly} universal \cite{sadhukhan2022origin} (Supplementary Figs. S12).
\begin{equation}
 \label{eq:1}
   P(x, k)=\left[k^k / \Gamma(k)\right] x^{k-1} \exp [-k x]
\end{equation}
\begin{equation}
\label{eq:2}
    P(AR)=\frac{1}{\mathcal{N}}\left(AR+\frac{1}{AR}\right)^{3 / 2}\left(1-\frac{1}{AR^2}\right) e^{-\alpha\left(AR+\frac{1}{AR}\right)},
\end{equation}
In the first expression, $\Gamma(k)$, is the Legendre gamma function; in the second, $N$, is the normalization constant, and $\alpha$ is a system-specific parameter predicted to correlate with dynamics. The PDF of $x$ obtained from \eqref{eq:2} becomes nearly independent of $\alpha$. For all values of $\phi$ close to confluence and all $\tau_p$ values, $k$ hovers between 2.5$-$2.8 like that observed in confluent epithelia \cite{atia2018geometric}.

Equation \eqref{eq:2} follows from a mean-field theory \cite{sadhukhan2022origin}, which also anticipates that for a confluent monolayer of closed-loop objects, which applies to both living and our synthetic cells, $SD(AR) = 0.71\overline{AR} - 0.75$ (black line in Fig. \ref{Figure3}a). The agreement between our experiments and theory is excellent even for densities far from confluence ($\phi = 0.74$) (inset to (Fig. \ref{Figure3}a)), although the agreement becomes weaker for smaller $\phi$ values (Supplementary Fig. S8). Notably, we also find that $\alpha$ scales linearly with $\text{log}(\tau_\alpha)$, as predicted \cite{sadhukhan2022origin}, indicating a strong link between structure and dynamics (Supplementary Fig. S13).

\subsection{Dynamical heterogeneities, cell shape and shape variability}
Although it is widely recognized that the dynamics of dense passive/active liquids and glasses is spatiotemporally heterogeneous and consist of domains with different relaxation times \cite{berthier2011theoretical,gokhale2016deconstructing,berthier2014prl,arora2022prl}, the correlation between cell aspect ratio and dynamics, as we report here so far, and like in previous studies \cite{bi2015density,park2015unjamming,atia2018geometric}, has been arrived at by system-wide averaging. In fact, recent experiments on confluent cell monolayers and 3D tissues show that such a correlation is also present locally, i.e., cells in the faster relaxing regions are more elongated than those in the slower relaxing ones \cite{kas2021frontiers,grosser2021prx}. Additionally, since at the system level, $SD(AR)$ and $\overline{AR}$ are mutually constrained (Fig. \ref{Figure3}a), it is reasonable to expect that cell shape variability in these faster regions is also correspondingly larger, although this remains untested.

\begin{figure}[tbp]
\includegraphics[width=0.95\textwidth]{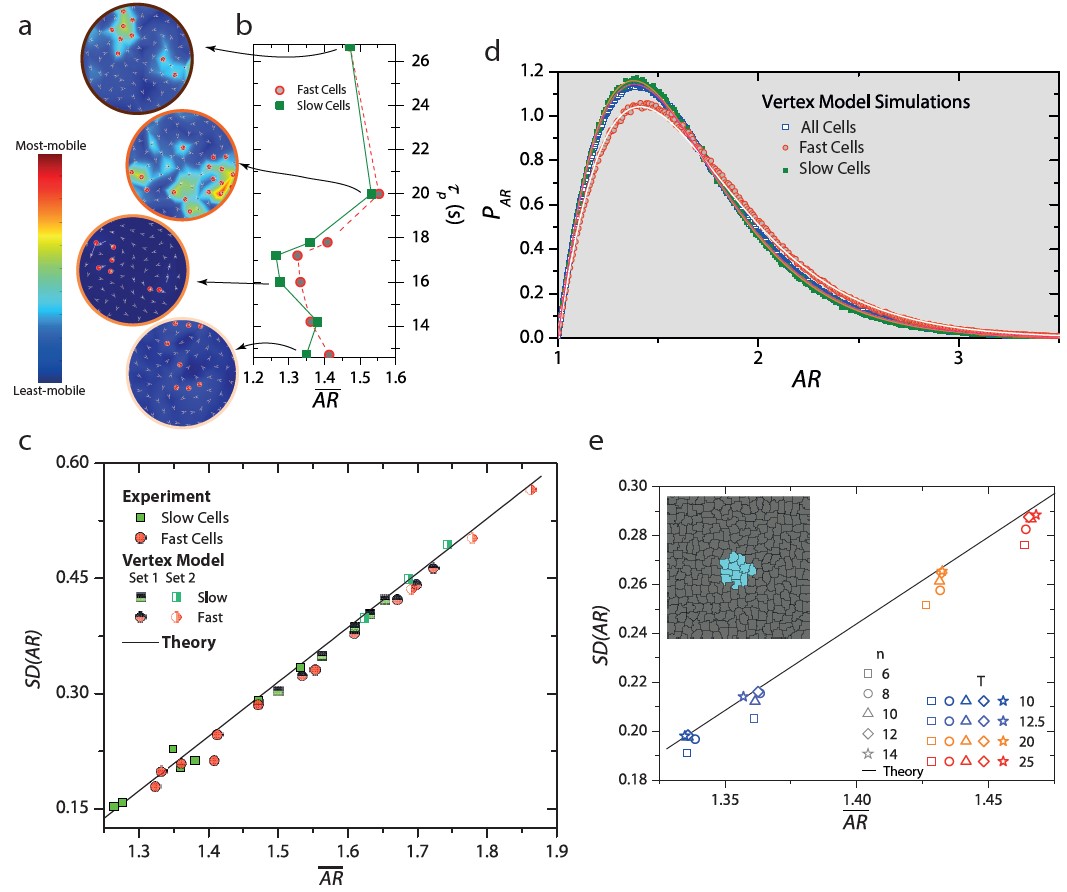}
\caption{\textbf{Fast cells show suppressed shape variability.} 
\textbf{(a)} Cell velocity maps over the cage breaking time, $t^\star$, for different $\tau_p$ values for $\phi=0.92$. The color bar denotes the displacement magnitude. The red circles represent the top 10 $\%$  most-mobile cells over $t^\star$.
\textbf{(b)} $\overline{AR}$ for different values of $\tau_p$ for the top 10 $\%$  most-mobile (hollow circles) and least-mobile cells (squares) cells over $t^\star$. Fast cells typically have a larger $\overline{AR}$ than the slow cells.
\textbf{(c)} $SD(AR)$ versus $\overline{AR}$ for the fast (circles) and slow cells (squares). The fast cells show a reduced shape variability in both experiments (filled symbols) and Vertex Model simulations (half-filled symbols).
\textbf{(d)} PDF of $AR$ for all cells (hollow squares), 10 $\%$ fast (red circles) and 10 $\%$ slow cells (solid squares) from Vertex model simulations at $T = 0.009$. \textbf{(e)} Inset: Mimicking dynamical heterogeneities. After equilibrating the system, a cluster of $n$ cells (shown in cyan) is allowed to evolve while the rest of the system is frozen (gray). Main figure: $SD(AR)$ versus $\overline{AR}$ for different $n$ and at different $T$. The black line is the universal scaling predicted in ref. \cite{sadhukhan2022origin}. With increasing $n$, $SD(AR)$ approaches the universal scaling from below, demonstrating that the subdued shape variability of fast cells is due to their confinement by their slower surrounding. Also, at low $T$, where the mobility of the system as a whole is smaller, the deviation from the line is also small. At high $T$, on the other hand, the confinement effect is more pronounced.
}
\label{Figure4}
\end{figure}

To this end, we followed standard procedures to quantify dynamical heterogeneities \cite{berthier2011theoretical,gokhale2016deconstructing} and first identified the top $10$\% fastest and slowest cells over the cage-rearrangement time, $t^*$ (Supplementary Fig. S14). In  Fig. \ref{Figure4}a, we show the fastest cells (red circles) overlaid on the cell displacement maps also generated over $t^*$ for different $\tau_p$ values for $\phi = 0.92$. As is typical of deeply supercooled liquids, these cells are spatially clustered, which is the case with slow cells as well, and further, these fast cells within a cluster move together largely as a flock. Also, the heterogeneity is most pronounced at an intermediate value of $\tau_p$ (Supplementary Movie S7). Next, we determined the mean and standard deviation of the aspect ratio of the cells within these fast and slow clusters. Notably, fast cells have a larger $\overline{AR}$ than slow cells (Fig. \ref{Figure4}b). However, while for the slow cells, the growth of $SD(AR)$ with $\overline{AR}$ follows the universal scaling, for the fast cells, surprisingly, it lies below the predicted line, implying that these cells show reduced shape variability (solid symbols in Fig. \ref{Figure4}c).

To strengthen these findings and examine their generality, we performed simulations of the Vertex model (VM) \cite{farhadifar2007currbio} and the Cellular Potts model (CPM) \cite{noppe2015intbio}, which are highly effective representations of confluent epithelia. The larger system size and superior temporal statistics of the simulations also allowed quantifying the $PDF$ for the fast and slow cells. We observed that while the $PDF$ of the slow cells overlapped with that of the entire system, the $PDF$ for the fast cells had a larger skew, as is expected for a locally unjammed region of the system (Fig. \ref{Figure4}d and Supplementary Fig. S15). It is noteworthy that even in the simulations, the growth of $SD(AR)$ with $\overline{AR}$ for slow cells followed the predicted scaling, while for fast cells, $SD(AR)$ was lower than anticipated (half-filled symbols in Fig. \ref{Figure4}c and Supplementary Fig. S16).

The fact that most-mobile cells do not conform to the universal scaling is, in all likelihood, because the scaling comes from a mean-field theory, and hence, fluctuations are ignored. Mean-field approaches fail to capture the physics of dynamical heterogeneities even in conventional dense liquids and glasses \cite{berthier2011theoretical}. Nonetheless, to gain insights into our observations, we made the simplifying assumption that on a time scale comparable to $t^*$, fast cells are essentially confined within a frozen exterior of slower cells due to their different relaxation rates. To mimic this effect numerically, we first allowed the entire system to equilibrate over many times $\tau_\alpha$, and then froze all but a cluster of $n$ cells in the system interior (shown in cyan in the inset to Fig. \ref{Figure4}e). We performed simulations at different temperatures to let the system access different values of $\overline{AR}$. Trivially, when all but one cell is frozen ($n = 1$), there can be no shape variability, and $SD(AR) = 0$. Indeed, on increasing $n$, $SD(AR)$ increased systematically and followed the predicted scaling for $n\geq12$. This observation nicely demonstrates that the subdued shape variability of the fast cells is an outcome of the temporary confinement imposed by their slower neighbors.  

\subsection{Discussion}
Our confluent synthetic cell monolayers are an oversimplification of confluent epithelia. Unlike in the latter, no explicit cell-cell adhesion exists, although an effective activity-mediated attraction emerges here.   Also, our cells have a fixed perimeter, which is not the case with live cells. The area constraint is weaker in our system than confluent epithelia, but it still exists because of a dynamic internal skeleton. Nonetheless, in the context of jamming, our synthetic cell collective behaves like a living one, thus proving that geometric constraints alone can govern the dynamics of deformable active matter systems, be they inert or living. The exquisite control offered by our synthetic model system allowed us to discover a re-entrant jamming transition mediated solely by cell shape changes. Additionally, by dissecting dynamical heterogeneities, we have shown that transient confinement of the fast cells by their slower neighbors suppresses shape variability and leads to the violation of mean-field predictions. Cell shape, on the other hand, correlates with both global and local dynamics.


On the experimental front, the model system developed here allows for precise and easy tuning of many critical physical parameters like the chirality of the assembly, individual cell membrane stiffness, the nature of activity within the membranes, and even the inter-cell friction. Although each of these parameters has been identified to substantially impact the form and function of the cell collective \cite{sato2015prl,fuhs2022rigid,angelini2012fd}, these are impossible to control systematically in real systems, and some of these, like inter-cell friction, are yet to be incorporated even within the models of confluent epithelia. Through our preliminary experiments with magnetic membranes, we can already modify individual cell properties in real time. The experimental advance made here now makes it possible to dissect the role of heterogeneous cell properties on collective behavior \cite{li2019prl} and in fundamental processes like cell sorting \cite{lawson2021curropncellbio} from a purely geometric/mechanical perspective \cite{manning2023essay, graner2017devlopment,thomson2017book}.

\section{Author Contributions}
\textbf{P.A.}: Conceptualization, Methodology (experiments), Software, Validation, Formal Analysis, Investigation, Data Curation, Visualization, Writing - Original Draft, Writing - Review \& Editing. \textbf{S.S.}: Methodology (simulations), Software, Validation, Formal Analysis, Data Curation, Writing - Review \& Editing. \textbf{S.K.N.}: Methodology (simulations), Software, Validation, Data Curation, Formal Analysis, Writing - Review \& Editing, Supervision (simulations), Funding Acquisition. \textbf{D.B.}: Formal Analysis, Validation, Writing - Review \& Editing. \textbf{A.K.S.}: Validation, Writing - Review \& Editing. \textbf{R.G.}: Conceptualization, Methodology (experiments), Validation, Investigation, Formal Analysis, Visualization, Writing - Original Draft, Writing - Review \& Editing, Supervision (experiments), Project Administration, Funding Acquisition.

\section{Acknowledgements}
\textbf{P.A.} thanks the Jawaharlal Nehru Centre for Advanced Scientific Research, Bangalore, INDIA for a research fellowship. \textbf{S.S} and \textbf{S.K.N.} acknowledge the support of the Department of Atomic Energy, Government of India, under Project Identification No. RTI 4007. SKN thanks SERB for grant via SRG/2021/002014. \textbf{A.K.S.} thanks the Science and Engineering Research Board, Government of India for the National Science Chair 
. \textbf{R.G.} thanks the Department of Science and Technology, INDIA, for financial support through the SwarnaJayanti Fellowship Grant (DST/SJF/PSA-03/2017-22).

\section{Data Availability}
All relevant data are available from the corresponding author upon reasonable request.

\clearpage

\end{document}